\documentclass[12pt]{article}
\usepackage{amsmath}
\usepackage{graphicx,psfrag,epsf}
\usepackage{enumerate}
\usepackage{natbib}
\usepackage{url} 

\usepackage{csquotes}
\usepackage{rotating}
\usepackage{pifont}
\let\proglang=\textsf

\newcommand{\blind}{0}

\begin{document}

\def\spacingset#1{\renewcommand{\baselinestretch}%
{#1}\small\normalsize} \spacingset{1}


\if0\blind
{
 \title{\bf Key attributes of a modern statistical computing tool}
 \author{Amelia McNamara \thanks{The author is grateful to Nick Horton and Rob Gould, who read drafts of this paper and gave many insightful comments. Thanks also to Alan Kay and Mark Hansen, who read early versions of this work in my doctoral dissertation.  }
 \hspace{.2cm}\\
  Statistical and Data Sciences, Smith College}
 \maketitle
} \fi

\if1\blind
{
 \bigskip
 \bigskip
 \bigskip
 \begin{center}
  {\LARGE\bf Key attributes of a modern statistical computing tool}
\end{center}
 \medskip
} \fi

\bigskip
\begin{abstract}
In the 1990s, statisticians began thinking in a principled way about how computation could better support the learning and doing of statistics. Since then, the pace of software development has accelerated, advancements in computing and data science have moved the goalposts, and it is time to reassess. Software continues to be developed to help do and learn statistics, but there is little critical evaluation of the resulting tools, and no accepted framework with which to critique them. This paper presents a set of attributes necessary for a modern statistical computing tool. The framework was designed to be broadly applicable to both novice and expert users, with a particular focus on making more supportive statistical computing environments. 

A modern statistical computing tool should be accessible, provide easy entry, privilege data as a first-order object, support exploratory and confirmatory analysis, allow for flexible plot creation, support randomization, be interactive, include inherent documentation, support narrative, publishing, and reproducibility, and be flexible to extensions. Ideally, all these attributes could be incorporated into one tool, supporting users at all levels, but a more reasonable goal is for tools designed for novices and professionals to `reach across the gap,' taking inspiration from each others' strengths. 
\end{abstract}

\noindent%
\vfill

\newpage
\spacingset{1.45} 

\section*{INTRODUCTION}
\label{sec:intro}
Tools shape the way we see the world, and statistical computing tools are no exception. Affordances for building graphics, representing data, and modifying analysis all impact how users conceive of their statistical products. As our world becomes increasingly data-driven, it is important to critically examine the tools we are using and look toward the future of computational possibilities. 

The use of the term `tool' to mean computer software or a programming language harkens to a time when computers do more than just amplify human abilities: they also augment them~\citep{Pea1985}. In the same way physical tools allow us to do more than we could on our own, computers can allow humans to `see'  and `think with' data in higher dimensions and with more clarity than they otherwise could. 

Statistical computing tools have historically been delineated into tools for learning and tools for doing statistics~\citep{Bag2013, McN2015}. Interestingly, while statisticians have thought and written about principles underlying the tools for learning statistics, almost no critical work has been done to evaluate professional tools for doing statistics. In the educational context, Rolf Biehler's 1997 paper Software for Learning and for Doing Statistics outlined principles for a statistical computing tool that would support novices in learning statistics and data analysis~\citep{Bie1997}. It provided a framework for the assessment of statistical education software, and motivated the development of new tools. The motivation and criticism of professional tools is much less rich, and tends to focus on language properties~\citep{IhaGen1996, MorHilOsv2012}. Today, the lines between educational and professional tools are starting to blur, and we believe this lowers the barrier to entry for statistical computing. 

This paper presents a list of attributes capturing features needed for tools for both novices and professionals. The attributes aim to be as broad as possible, but they are focused specifically on the development of more supportive environments. Hopefully, this list can be used to frame critical discussions of statistical computing tools of all types. The target audience of this paper are software developers who are considering the development and improvement of statistical computing tools. Practicing educators may also find the paper interesting, as it forms a scaffolding for evaluating existing tools and deciding which to use in a particular course. And, statisticians could do to think more critically about the tools they use and help create. 

Although the attributes have been designed to be applicable to a broad range of users, it is useful to focus our discussion on ``a user'' rather than ``the user''~\citep{Agr1995}. For purposes of discussion, we consider our target user to be a journalist looking to bring more computation into their work. The practice of data journalism has been accelerating, but journalism schools have been slow to modify the curriculum to teach computational skills~\citep{BerPhi2016},  As a result, many journalists have limited experience with programming and statistics, but want to tell data-driven stories. They need to move from novices to producers very quickly. A key goal of journalism is communication, and as news publications have embraced the interactive web, journalists are at on the forefront of publishing modern, data-rich reports. Considering a data journalist as our target user means prioritizing tools that are easy to learn but also powerful and flexible. 

We could have focused just as easily on a number of other specific users. For example, a graduate student in a scientific or social scientific field who needs to use statistical methodology to complete their work. We may imagine those graduate students are already using a statistical computing tool, but many use Excel~\citep{Wei2017}, and if they need to get up to speed with another package they are often self-taught~\citep{Low2017}. Once again, these users need tools that are easy to use, and become `invisible,' allowing them to get their work done. Because of the increasing importance of reproducibility in science, they also need tools that allow them to document their work. For someone who uses a statistical computing package every day, it may be hard to imagine being new to it, or conceptualize how the tool could be improved. However, in order to broaden the use and understanding of statistics, we need to make it easier to do statistics, and to do statistics well. 

Many of the ideas presented here are not new. In particular, these attributes attempt to distill principles and characteristics proposed by Rolf Bieher, Alexander Repenning, and John Tukey~\citep{Bie1997, RepWeb2010,Tuk1965}, while also considering the recent developments in data and computing. John Tukey was considering the ``technical tools of statistics'' in 1965, and describing a vision for the future of statistical programming tools, including `More of the essential erector-set character of data-analysis techniques, in which a kit of pieces are available for assembly into any of a multitude of analytical schemes''~\citep{Tuk1965}. Thirty years later, Rolf Biehler defined three primary problems, the ``complexity of tool problem'' (existing tools were too hard to for novices to learn), the ``closed microworld problem`` (learning tools were designed for one particular type of problem or data set and couldn't be extended) and the ``variety problem'' (because of the closed microworld problem, it was necessary to use many tools to do everything an instructor wanted to cover)~\citep{Bie1997}. Most recently, Alexander Repenning, David Webb, and Andri Ioannidou outlined the six requirements for a ``computational thinking tool,'' including having a low threshold, a high ceiling, and being equitable~\citep{RepWeb2010}.

A survey of statistical computing tools~\citep{McN2016b} helps ground these ideas in the existing computational landscape. Again, since statistical computing tools have often been delineated into professional and educational tools, we take representative examples from each `category' when we refer to existing software. When tools for learning statistics are mentioned in this paper, the most common examples will be TinkerPlots and Fathom, two interactive tools for statistics education~\citep{KonMil2005, Fin2002}. These tools are graphical, drag-and-drop interfaces that make analysis highly visual. Most references to professional tools will be to the programming language \proglang{R}~\citep{RCore2016} or to SAS, Stata, and SPSS. 

Considering the various positive qualities of current tools for doing and teaching statistics alongside Biehler's goals~\citep{Bie1997} and combining them with ideas from \cite{RepWeb2010} and \cite{Tuk1965}, we developed a list of 10 attributes for a modern statistical computing tool. These are summarized in Table \ref{attributelist}. While these attributes attempt to be exhaustive, they are also designed as a springboard for discussion. Because there has been relatively little recent critique given to statistical computing tools, this list is an attempt to start the conversation. 

\begin{table}[htbp]
\centering
\begin{tabular}{|l|}
\hline 
1. Accessibility \\
2. Easy entry for novice users \\
3. Data as a first-order persistent object \\
4. Support for a cycle of exploratory and confirmatory analysis \\
5. Flexible plot creation \\
6. Support for randomization throughout \\
7. Interactivity at every level \\
8. Inherent documentation \\
9. Simple support for narrative, publishing, and reproducibility \\
10. Flexibility to build extensions \\
\hline
\end{tabular}
\caption{Summary of attributes \label{attributelist}}
\end{table}

\noindent Each requirement will be discussed in more detail in its respective section. 
\clearpage

\section{Accessibility}
Tools should always be accessible, particularly to new users. As a baseline requirement, software should be affordable (or free), work with a variety of operating systems, and be easy to install~\citep{DunHen2008, RepWeb2010}. In this context, most tools for teaching are accessible, because they are designed to work across platforms and are priced inexpensively. However, professional tools tend to be expensive and inaccessible for non-professional or occasional users. They are not accessible for small newspapers, nonprofits, individuals, or K-12 school systems.

Users in these contexts must consider if they have the funding for a tool, if it will work on the computers they have access to, and if they have the user privileges to install software. System administrators can be few and far between in newsrooms and underfunded school systems. One way to ensure accessibility is to create a web-based tool---cloud services allow users to access software from any machine with internet access. 

Beyond the accessibility of a tool to the masses, it is also important to consider the needs of people with disabilities. For a tool to be required in public schools, it must be compatible with accessibility features on modern computers~\citep{DeptEd2001}. Some progress has been made on programming languages accessible for blind users~\citep{SteHun2011, God2013}, but given that many educational tools are visual, it is not clear if any of them are accessible to blind users. Of course, there are other disabilities that can impact a person's ability to use a tool. Considering ``universal design'' (a principle of designing things to be usable by all people)~\citep{ConJon1997} should be an aspect of the development of any new statistical computing tool. 

\section{Easy Entry For Novice Users}
Tools to be used by novices---and really, all tools---should make it easy to get started. This attribute comes directly from Reppenning et al's work on tools for computational thinking~\citep{RepWeb2010}. It should be clear what the tool does, how to use it, and what the most salient components are. The tool should provide immediate gratification, rather than a period of frustration eventually leading to success. 

Easy entry means users should be able to jump directly into `doing' data analysis without having to think about minutiae. Novices should to be able to begin exploratory data analysis within the first 10-15 minutes of using a tool. 

Biehler argued, ``In secondary education, but also in introductory statistics in higher education, using a command language system is problematical. We are convinced that a host system with a graphical user interface offers a more adequate basis''~\citep{Bie1997}. By Biehler's estimation, a system providing easy entry for novices will likely have a visual component, either initially or throughout.  

Indeed, visual tools like TinkerPlots and Fathom allow novices to create linked plots and multivariate data visualizations within the first minute of beginning the software. Curriculum development using the programming language \proglang{R} has begun to put first emphasis on exploratory data analysis, rather than data structures~\citep{PruHor2014}, so these goals can also be achieved in a scripting context. Given the success of the blocks-based language Scratch in computer science education~\citep{ResMalMon2009}, it seems possible a graphical system would be better for novices. However, there are many other ways easy entry could be achieved, such as the use of language levels~\citep{HsiSim2005}, or accessible IDEs~\citep{Kol2010}. 

\section{Data as a First-Order Persistent Object}
A data analysis tool must necessarily deal with data. A tool cannot be considered to be designed for statistical computing if it does not make data its primary object of interest. The way data are formatted and represented within the system is also of crucial importance. In this context, formatting and representation refer specifically to how the data appear to the user, not how they are stored within the computer's memory system. 

Modern data analysis tools should make it easy to access common data types (flat files, hierarchical data formats, APIs, etc.) and `see' the full data (whether in a format allowing for value-reading or a higher-level view). Data should be a persistent object, with a reproducible workflow of wrangling to take raw data to clean.

\subsection{Viewing data}
Many tools (including spreadsheets) make a view of the data the primary focus. In conversations with data journalists, they often mention scrolling through a spreadsheet, `reading' the data values as their first line of inquiry. Scientists also like to visually look through their data when they begin. While there are few recent studies to support this, an early experiment on Lotus 1-2-3 suggested users spend around 42\% of their time viewing individual cells~\citep{BroGou1986}. In contrast, programming languages like \proglang{R} and \proglang{Python} have traditionally not shown data to users when it is read in, instead requiring the use of function calls to view the first few rows of data. This can be a sticking point for users transitioning from spreadsheet programs, so Integrated Development Environments providing a data preview have become popular~\citep{RStudio2014}. 

Whether provided by default or requested by the user, most current tools provide data such that users can immediately read each individual value. However, there are other ways to `see' an entire data set. For example, Victor Powell's CSV Fingerprint creates a colored image as a high-level overview of the data~\citep{Pow2014}. Colors indicate data types (to see whether it is mostly numeric, categorical, integer, etc) and missing data~\citep{Pow2014}. This simple visualization gives a lot of insight, and suggests that there may be other visual metaphors to represent raw data that could be equally helpful. The more a user can glean from an initial glance at their data, the easier it is for them to begin to dig in to it. 

\subsection{Rectangular versus hierarchical data}
Analysts are typically accustomed to thinking of data in a tidy rectangular format, composed of rows and columns or observations and variables. Rectangular data can be considered `tidy' if every row represents one case (e.g., a person, gene expression, or experiment), and every column represents a variable (i.e., something measured or recorded about the case)~\citep{Wic2014}. Tidy data are often visualized as a spreadsheet, and spreadsheets are the most common way people around the world interact with data~\citep{Bry2016}. 

Interestingly, novices who have not encountered data before often do not use rectangular formats to represent their data, but rather default to a list-based or hierarchical format~\citep{LehSch2007, Fin2013, Fin2014}. So, although rectangular data has become prevalent, it may not be the most natural format. There are hierarchical and list-based data formats like \proglang{JSON} and \proglang{XML} which are used commonly on the web. These types of data are important for data science~\citep{NoTem2014}. However, they may require the development of new visual metaphors because the tidy rectangle will no longer suffice. How can you see a clear overview of an entire hierarchical dataset? One observation may stretch down the screen, and Powell's colored overview certainly does not directly translate.

\section{Support for a Cycle of Exploratory and Confirmatory Analysis}

Statistical computing tools should promote exploratory analysis, and its twin, confirmatory analysis. The complementary exploratory and confirmatory cycles were suggested by John Tukey in his 1977 book, and have been re-emphasized by current educators~\citep{Tuk1977, Wei2005, BieBenBak2013}. The use of the term `cycle' indicates how iterative the data analysis process is. Each step can lead back to prior steps.  The cycle can include generating statistical questions, collecting data, analyzing data, and interpreting results~\citep{CarEve2016}. Hadley Wickham lists import, tidy, transform, visualize, model, communicate~\citep{Wic2014c}.  In a pedagogical setting, educators often talk about the PPDAC cycle: Problem, Plan, Data, Analysis, Conclusions, typically attributed to \cite{WilPfa1999}. 

If users find something interesting in a cycle of exploratory analysis, they need to follow with confirmatory analysis. The idea of EDA is to explore data deeply by computing descriptive statistics and making many graphs---of one variable or several-- to gain an understanding of the underlying structure. Although EDA can appear subjective, it sometimes comprises the best and richest method for analysis, particularly for finding patterns in data and performing informal inference~\citep{MakRub2014, RubHamKon2006}. Exploratory data analysis can also be used in the context of statistical modeling~\citep{Gel2004}.  

The difference between exploratory and confirmatory analysis (or informal and formal inference) is like the difference between sketching or taking notes and the act of creating the final painting or writing an essay. One is more creative and expansive, and the other tries to pin down the particular information to be highlighted in the final product. A system supporting exploration and confirmation should provide a workflow connecting these two types of activities. Users need `scratch paper'---a place to play without the results being set in stone. While data analysis needs to leave a clear trail of what was done so someone else can reproduce it, a scratch paper environment might allow a user to perform actions not `allowed' in the final product, like moving data points. Biehler called this capability `draft results'~\citep{Bie1997}. 

Many current systems for teaching statistics provide rapid exploration and prototyping (allowing users to manipulate data or play with graphic representations), but typically do not support the more formal final analysis. In contrast, professional tools tend to make it difficult to play with data (in \proglang{R}, creating multiple graphs takes effort, as does modifying parameter values), and they may not support cyclical exploration or rapid plot generation. Again, this is limiting, as a sense of play and discovery is important to data analysis. Data scientists repeatedly cycle back through questioning, exploration, and confirmation or inference, so analysis is never a linear process from beginning to end. A statistical computation tool should support this cyclical process.

\section{Flexible Plot Creation}\label{plots}
To fully support data analysis (both exploratory and confirmatory), a tool needs to emphasize plotting. Computational tools make it possible to visually explore large datasets in ways that would be difficult or impossible using just pencil and paper. Visualization greats John Tukey and Jacques Bertin both developed visualization methods for summarizing and visualizing patterns in data before computer graphics~\citep{Tuk1977, Ber1983}. 

These static plots are still useful now that computers can generate them, but a statistical computing tool should give humans abilities beyond what they could achieve with pencil and paper. An exemplary method is the Grand Tour, which takes high-dimensional data and produces projections into a variety of 2- and 3-D spaces, walking a user through many views of their data to expose clusters and trends~\citep{BujAsi1986}. A simpler example that can also provide insight is the generalized pairs plot, which displays all 2-variable relationships in the data~\citep{EmeGreSch2013}. These plots allow humans to look for patterns in higher dimensions than they could ordinarily conceptualize.

Providing easy plotting functionality of many variables should be a goal of every tool, whether for learning or for doing statistics. Tools, particularly those for novices, must choose whether to provide a few simple plotting functions or the ability to fully customize graphics. While it can seem simpler to provide a small set of standard data visualizations, creating visualizations from primitives both provides more flexibility for the user and reinforces the mapping between abstract data and visual aesthetics on the screen~\citep{Wei2005, Wil2005, Wic2009}. Ideally, a statistical programming tool would make it simple to begin plotting (to facilitate EDA) and to produce standard graphics, while also allowing users to create novel plot types.

\section{Support for Randomization Throughout}\label{randomization}

Computers have made it possible to use randomization and bootstrap methods where approximating formulas would once have been the only recourse. These methods are not only more flexible than traditional statistical tests, but can also be more intuitive for novices to understand~\citep{PfaWil2014, TinTop2012}. Randomization and simulation can help make inference from data, even if those data are from small sample sizes or non-random collection methods~\citep{EfrTib1986, Lun1999, Ern2004}.  

Randomization and the bootstrap can also be used to validate models~\citep{MajHof2013, BujCooHof2009, Gel2004}, provide a visual representation of uncertainty in a plot~\citep{HulRes2015}, or perform graphical inference, a method of assigning significance to plots by using a series of randomized plots to provide a ``null'' visualizations to compare true visualizations against~\citep{Wic2010,MajHof2013, BujCooHof2009}. 

These methods have been gaining popularity in statistics research and trickling down to the educational context as well. Several popular introductory statistics textbooks focus on randomization and simulation methods~\citep{DieBar2014, TinCha2014, Lock5}, and other resources help get instructors up to speed~\citep{Hes2014}. These materials avoid the issue that many introductory statistics courses fall into, where the course can begin to feel like a grab-bag of methods. Instead, they show randomization as a unifying method to answer many statistical questions using one framework. 

The application of randomization and the bootstrap is a place where tools for teaching statistics shine. Popular applet collections provide simple randomization and bootstrap functionality~\citep{ChaRos2006, LocLoc2014}. TinkerPlots and Fathom also provide intuitive visual interfaces for this~\citep{Fin2002, KonMil2005}. However, professional tools have lagged behind. \proglang{R} provides the most complete functionality, but it is not always simple to use. 

Because of their intuitive nature and generalizability, randomization and bootstrap methods can be helpful for novices and experts alike. They can be used in a variety of contexts, including graphical inference methods bridging the gap between exploratory and confirmatory analysis. 

\section{Interactivity at Every Level}\label{interactivity}
Interactive systems enable users to be more engaged and playful with data. Rather than typing commands, users should be able to interact with their data. And the more direct the manipulation, the better. This means valuing pinch-zoom over a dropdown menu with an option for zoom, click-and-drag selection over a form allowing the user to enter filtering values, and linked plots and analysis over a set of disconnected products. Here, products encompasses anything that comes out of the analysis, including plots, model output, and summary statistics. 

Interactivity is becoming standard on the web. Users of Google maps know they can pan and zoom a map, and Apple has strong opinions on which direction is more `natural' to scroll. On smartphones we launch angry birds, drop pins on our location, and swipe left to reject a date. 

Data analysis platforms need to follow suit. For novices, we want to ``Teach about, and with, interactive graphics''~\citep{Rid2015} so they become adept at seeing data in this way. As Biehler suggests, we want to encourage direct manipulation rather than modifying a script~\citep{Bie1997}. Today, educational tools provide this type of direct manipulation, but professional programming tools often do not. However, even textual programs can shorten the time between making a change in the code and seeing the results. Computer futurist Bret Victor has made shortening this loop one of his driving design principles, to provide users with the ability to see the direct results of their actions without waiting for something to compile~\citep{Vic2012}. The development of d3.express shows promise in bringing this paradigm to the visualization library d3~\citep{Bos2017}. 

In the context of statistical programming, Deborah Nolan and Duncan Temple Lang make the distinction between dynamic documents (those that are compiled and then automatically include the results of embedded code), and interactive documents (those that let a reader interact with components like graphics)~\citep{NolTem2007}. Given the goals of interactivity at every level, and the importance of publishing, a modern statistical programming tool should provide `dynamic-interactive' graphics, where users can interact with any component of the document and have the results update in real time. 

Interactivity can take place at three levels. The first is in the context of developing an analysis. Ideally, users should be able to build their analysis interactively. Menus and wizards are a type of `interaction,' but are not direct interaction and don't add any intuition about the process. Instead, a tool should aim to allow for the most direct manipulation possible. 

The second level is within the analysis session, where all results should themselves be interactive. The tool should support graphs as an interface to the data~\citep{Bie1997}. Behaviors like brushing and linking should do dynamic subsetting~\citep{Wil2005, Few2010}.  All graphs should be zoomable, support brushing and linking, and allow for simple tooltips to identify data points. It should be easy to change the data cleaning methods and see how that change is reflected in the analysis afterward, and parameters should be easily manipulable. The system should also make it possible to see multiple coordinated views of everything in the user's environment. The importance of a coordinated view is supported by researchers who suggest allowing for multiple views of the same data may help people gain a more intuitive understanding~\citep{ShaHoe2002, Bak2002}. 

Finally, the finished data product should be interactive. This means that the audience of a piece of data analysis---even if they do not know much about statistics---could play with the parameters and convince themselves the data were not doctored. 

As may be expected, standalone educational tools do a better job of providing interactivity than professional tools. 

TinkerPlots and Fathom are highly interactive, allowing users to drag-and-drop variables onto their plots and supporting brushing and linking between plots. Highlighting cases in the data table highlights them in every plot. These tools make it easy to interactively develop analysis and play with it, but do not support sharing interactive results with someone who does not have the software.

On the other hand, interaction has historically been more challenging in professional tools. The history of statistical computing traces back to the pre-graphics era of computers, so most systems rely on static code. This paradigm means users are not incentivized to return to the beginning of their analysis to see how a code modification would trickle down. If a programmer wants to adjust a parameter value in their code, they must modify the code and re-run it, making the comparison between states in their head.  Comparing two states in this way may be possible, but comparing more than two is difficult. This is a cognitive burden we no longer need to put on users~\citep{Vic2012}. If results were immediately accessible, it would make it possible to make hundreds of comparisons in just a few seconds.

In recent years, some of these possibilities have begun to emerge. `Notebook' functionality in several environments allows users to execute code chunks directly within their source file~\citep{PerGra2015,RStudio2016}. For experienced programmers, the production of interactive documents that respond to user input is possible~\citep{ChaChe2015, Bos2013, SatRusHofHee2016}. While these packages allow expert users to create dynamic graphics, they are too complicated for a beginner. 

As a result, most current published work with interactive abilities is the result of a bespoke process. Because few tools exist to facilitate the development of fully interactive data products, people who want to generate such products must hard-code them for a particular application. Two exemplary pieces of journalism include a simulation-based look at hurricane impacts in Houston by ProPublica, which allows readers to manipulate parameters of the simulation~\citep{Sat2016}, and the IEEE programming language ratings~\citep{CasDiaRom2014} which provides access to the weight parameters used for each data source in the rating algorithm. 

The power and usefulness of a truly interactive data analysis platform is easy to imagine. If all parameters were adjustable, it would be easier to get an intuitive sense of the parameter space, and therefore the fragility of a particular piece of an analysis. 

\section{Inherent Documentation}\label{visdocs}
Systems should provide inherent documentation, so computing tools ``highlight the logic of what is going on''~\citep{Kap2007}.  Most programming language documentation is hard for novices to comprehend, so we first want help that is helpful. However, the idea of a inherent documentation goes one step further, to help that is integrated into the process of using a tool. Instead of having to go to a second place to learn what a feature is or what a function does, objects should provide documentation as a unified part of themselves. 

Ideally, every component of a system should visually show the user what it is going to do, versus just telling them. However, even in textual languages inherent documentation can be achieved by bringing the syntax of the language more in line with human language. Function names that describe what they do are more valuable than those that preserve keystrokes. Supportive features like tab completion can make documentation of parameters more inherent to the analysis process. 

For example, if a tool is going to perform k-means clustering, the basic level of documentation should be the words ``k-means.'' Ideally, the user should see a visual representation of the algorithm, and as it is applied to the data, interim steps should be visualized~\citep{MuhPir2014}. Of course, using a computer is not the same as moving through the real world, so interface designers must think carefully about visual metaphors that make the most sense. Sometimes, this means mimicking the real world (as in the desktop metaphor, with folder icons and a trash can) and sometimes developing a new visual language (as may need to happen for visualization of models, database operations, and the like). Interactive controls of a system should give some idea of what they are going to do, either by their design or by the presentation of `scented widgets', embedded visualizations providing hints to users about what elements are capable of~\citep{PouStaMat2007}.

\section{Support for narrative, publishing, and reproducibility}\label{repro}

One important component of data science is the communication of results. We have already considered the importance of flexible plot creation, which is a form of visual communication. In addition to plots, almost all data analytic products require some form of narrative to accompany the work and contextualize it for readers. The products of a statistical computing system should be as easy to understand as the process of creating them, and they should be simple to share with others. Integrated narrative and button-click publishing will provide affordances that support reproducibility. Reproducible, interactive workflows may help to build confidence in results because they can be easily verified even by non-experts. 

\subsection{Narrative}
Historically, analysis workflows have tended toward a paradigm of doing analysis in one document and narrative in another. Programmers traditionally separate the documentation of their code from the code itself (code comments notwithstanding). Data analysts often create their data analysis code first, then go back to create a narrative surrounding the analysis. Data journalists refer to the process of performing analysis in Excel and writing about the results in Word as keeping a `data diary.' 

In contrast, a statistical programming tool should have affordances to encourage narrative alongside or mixed in with the code to facilitate the integration of storytelling and statistical products. Donald Knuth calls this `literate programming' because it is easier for humans to read and understand~\citep{Knu1984}. 

Currently, the most successful tools allow users to write formatted text and delimited code, then process the document to create a final product with text, code, and code output~\citep{PerGra2015, Xie2014}. Even those tools leave something to be desired. They feel constrained, and do not lend themselves to the type of expressive work that characterizes data science. Delimiting code chunks is a fairly lightweight process, but it does require some additional syntax. And including incidental numbers into narrative sentences can be tricky. A better solution would allow for explicit linking between code chunks (or, automatic detection of reactive connections), and the ability to drop any piece of an analysis into the text. 

\subsection{Publishing}
Ideally, data analysis results and related products could be published with ease. Journalists could create a data-driven website, citizen scientists could share insights in the data they helped create with their friends and family, and people working together across an organization (or across the globe) could stay up-to-date on their collaborators' contributions. In all these scenarios, the publishing format should allow for exploration (discussed in more depth in Section \ref{interactivity}). In fact, the ideal case would be a finished product allowing for full access to all the computation in the analysis. In this way, users could continue to explore the data, modify the analysis, and see the effects of their changes on the analysis and visualizations. 

As the expected user base for analysis publication is wide (encompassing both novices and experts) the language the analysis is written in should be the same as the language it is published in. Currently, it is often necessary to translate from one format to another to share analysis. For example, a data journalist using RMarkdown to document their analysis will need to format it after the fact using their newspaper's content management system. To achieve the goal of native publishing, it is likely new linkage pipelines will need to be developed in order to streamline these transitions. 

In data journalism, simple publishing abilities for fully interactive results of a data analysis could empower journalists to produce richer articles. Such articles could be accompanied by the reproducible code that produced them, allowing readers to audit the story. Similarly, as reproducibility becomes more valued in the academic community, data products are more often accompanied with fully reproducible code. If the code were interactive, it would widen the potential audience of the academic work.

\subsection{Reproducibility}
Reproducibility supports the aims of science, and should therefore be integrated with the work of data science~\citep{BucDon1995, SanNek2013, IncHatCum2012, DeL2009}. Teaching novices to use tools that support reproducibility can help ensure it becomes an integral part of their statistical and data workflow~\citep{CarEve2016}.  

There are many definitions of reproducibility. Here, we take a somewhat narrow view. A reproducible analysis is one that can be re-run (potentially years later, or by a different person) with the same data to produce exactly the same result. A slight extension to this is an analysis that can be re-run with a modified version of the original data to get analogous results~\citep{KanHeePla2011, SanNek2013, Bro2015}. For example, the initial analysis was done on 2016 data but needed to be run again on 2017 data, or the initial analysis used corrupted data that should be replaced by a corrected version.

It may sound simple to achieve this goal. However, in practice there are many factors that make it challenging. Software versions can change, package dependencies can get broken, and---most disruptive to the process---authors often do not manage to document their entire process. They may have done data cleaning outside the main software package (e.g., the bulk of the analysis was done in \proglang{R} but the author did data cleaning in Excel before the analysis), or run analysis steps without adding them to the code document. They may provide out-of-date code, or code with bugs that need to be addressed before it will run. These problems can be at least partially addressed with tooling. 

Integrated narrative and simple publishing will necessarily encourage reproducibility. If analysis developers are writing narrative as they write code, the results will be easier to interpret and more likely to be housed in the same place. If it is easy to publish this type of document, readers will have access to a richer version of the analysis than is typically shared. Therefore, the products of statistical computing tools should continue to become more reproducible. 

However, there is more work to be done before any statistical computing tool can be said to fully support the entire spectrum of reproducibility. 

A fundamental feature supporting reproducibility is the ability to save the data analysis process. Some teaching tools (e.g., applets) do not allow state to be saved in any way. In other systems, like Fathom and Excel, analysis is not reproducible because it was produced interactively. Even in 1997, Rolf Biehler was aware of this drawback to interactive systems; ``It may be considered a weakness of systems like Data Desk that the linkage structure is not explicitly documented as it is the case with explicit programming or if we had written the list of commands in an editor. An improvement would be if a list of commands or another representation of the linkage structure would be generated automatically''~\citep{Bie1997}. Most interactive tools do allow the user to save the environment that produced the product, but do not document the steps taken within the environment. An independent researcher could use the saved document to explore the analysis, but may not be able to discover the steps to produce the final product. These types of tools also make it impossible to re-run the analysis on slightly different data. 

Again, professional tools allowing for the integration of narrative and code are beginning to support some of these goals. Using R and RMarkdown, for example, users can now author entire analyses within a single document, fulfilling Broman's `everything with a script' and `turn scripts into reproducible reports'~\citep{Bro2015, Xie2014}. Some of these tools are simple enough to be integrated in introductory college statistics courses~\citep{BauCet2014}. However, even experts trying to implement reproducible workflows have found it difficult to fully document their process~\citep{FitzPen2014, GarKin2013}. For novices, full reproducibility is even more challenging~\citep{GarKin2013}. 

Future systems should therefore be designed in order to support reproducibility more fully. This may entail saving a version of the computer's state, tracking all `scratch work' alongside code put into a `final draft,' automatically recognizing dependencies on files, packages, and custom functions, and providing a visual representation of those dependencies to the user. This vision would move close to Nolan and Temple Lang's vision of dynamic, interactive documents~\citep{NolTem2007}.

\section{Flexibility To Build Extensions}\label{modularity}

Of course, a statistical computing tool must have statistical methods built into it. While these attributes have outlined elements that approach methods (such as graphics and randomization) they shy away from specifying any particular models or techniques. This is because statistics is always changing, so one of the most important attributes of a statistical computing tool is the ability to extend it. 

The flexibility to build extensions is necessary in order to prevent a tool from becoming obsolete. Users must be able to create new components of the system as methods are developed, computers improve, or scientific discoveries are made. To be a computational thinking tool, building extensions is a required feature such that the system has a ``high ceiling,'' preventing users from `aging out' or `experiencing out' of a system~\citep{RepWeb2010}. In a statistical computing tool, it should be possible to develop new visualization types and data processes from other modular pieces.

Professional tools can be looked to for inspiration, because they tend make it easier to create new components of the system using old ones. \proglang{R} even has a centralized repository where other users can easily find and import others' work~\citep{CRAN2015}.  Currently, the tools easiest for novices to use fail to provide a high ceiling, although Biehler argued that ``adaptability (including extensibility) is a central requirement for data analysis systems to cope with the variety of needs and users''~ \citep{Bie1997}. 

Any system hoping to stay the test of time must provide the flexibility to build extensions. 

\section{CONCLUSION}
This list of 10 attributes aims to encompass the most important qualities for a modern statistical computing tool. We have focused on an idealized data journalist as our target user, but hopefully the attributes are more broadly relevant, encompassing some of the needs of science and social-science graduate students, novices at a variety of other ages, and seasoned statistics professionals. 

Of course, there are other features that one might desire for their tools. The list focuses on things that could be built into a system by an engineer, which overlooks the importance of a welcoming and supportive community of users. It also has not touched on the language attributes commonly cited by computer scientists, such as speed and completeness, and it assumes tools would be stable and free of errors. Does the ideal tool need to support Bayesian statistics? Should it include an algebra solver? While some of these questions can be encompassed into the ``flexibility to build extensions,'' there are certainly open questions.  More than anything, this list of attributes was designed to start a critical conversation about the design of statistical computing tools. 

Considering the existing tools for statistical computing, ~\cite{McN2016b} suggests that none of them fulfill all the attributes outlined above. Most tools can be described as either a tool for learning statistics or a tool for doing statistics. Those for learning statistics tend to be better at accessibility, easy entry, exploratory data analysis, flexible plot creation, randomization, and interactivity. For example, TinkerPlots and Fathom are highly interactive and intuitive, but make it difficult to share results. Spreadsheets like Excel are highly accessible to a broad audience, but obscure the computational processes taking place. In contrast, professional tools like R privilege data as a first-order object, support reproducibility, and have the flexibility to build extensions, but are harder to get started using and the data-analytic products they create are usually not interactive. For more details, see \cite{McN2016b}. 

No existing tools currently satisfy all the attributes, which suggests the need for new or improved software. It would be ideal to conceive of a single tool that could support users at all levels. For example, a blocks programming language with streamlined domain-specific language could step novices into more complex analysis. However, there are few examples of similar tools in other domains so it seems unlikely such a system will emerge, and indeed, projects which try to be all things for all people often fail. 

If we acknowledge that users will likely have to move from one type of tool to another, software developers should be looking for ways to `bridge the gap' between the two types of tools~\citep{McN2015}. In other words, in tools with traditionally difficult learning curves, designers should consider how to lower the barrier to entry, while in tools where users tend to `experience out', designers should build (either technically or pedagogically) an onramp toward the next tool. \proglang{R} has historically been difficult to get started using, but curricula and packages have been developed to lower the barrier to entry~\citep{BauCet2014, PruHor2014}. Researchers have also begun studying instruction methods that best support learning of both statistics and statistical computing~\citep{Bag2013} These efforts have not solved the problem of easy entry, but are easing the transition. More work needs to be done, but other tools could take inspiration from these initial efforts. 

As new tools are developed and existing ones are refined, statistical practitioners need to remain actively engaged in their development and critique to ensure they can support learning as well as doing statistics. Hopefully, this paper can act as a guide as we begin to engage more fully with this conversation.

\bibliographystyle{apalike}

\bibliography{../../../ReadingLibrary}

\end{document}